\documentclass[pra,twocolumn,superscriptaddress,showpacs,nofootinbibfloatfix,amsmath,amsfonts,amssymb]{revtex4-1}%
\usepackage{amsmath,amsfonts,amssymb,color}
\usepackage{amsthm}
\usepackage{leftidx}
\usepackage{graphicx}
\usepackage{xcolor}
\usepackage{dcolumn}
\usepackage{bm}
\usepackage{epstopdf}
\usepackage{epsfig}
\usepackage{environ}
\usepackage{pdfcomment}

\usepackage{float}
\usepackage[T1]{fontenc}
\usepackage[latin9]{inputenc}
\usepackage{setspace}
\usepackage{esint}

\begin{document}
\preprint{APS/123-QED}
\title{Topological delocalization transitions and mobility edges
	in the nonreciprocal Maryland model}
\author{Longwen Zhou}
\email{zhoulw13@u.nus.edu}
\affiliation{%
	College of Physics and Optoelectronic Engineering,
	Ocean University of China, Qingdao, China 266100
}

\author{Yongjian Gu}
\email{guyj@ouc.edu.cn}
\affiliation{%
	College of Physics and Optoelectronic Engineering,
	Ocean University of China, Qingdao, China 266100
}

\date{\today}
\begin{abstract}
Non-Hermitian effects could trigger spectrum, localization and topological
phase transitions in quasiperiodic lattices. We propose a non-Hermitian extension of the Maryland model, which forms a paradigm in the study of localization and quantum chaos by
introducing asymmetry to its hopping amplitudes. The resulting nonreciprocal
Maryland model is found to possess a real-to-complex spectrum transition
at a finite amount of hopping asymmetry, through which it changes
from a localized phase to a mobility edge phase. Explicit expressions
of the complex energy dispersions, phase boundaries and mobility edges are found. 
A topological winding number is further
introduced to characterize the transition between different phases.
Our work introduces a unique type of non-Hermitian quasicrystal,
which admits exactly obtainable phase diagrams, mobility edges, and
holding no extended phases at finite nonreciprocity in the thermodynamic limit.
\end{abstract}
\maketitle

\section{Introduction\label{sec:Int}}
Non-Hermitian topological matter has attracted great attention in
the past decade~(see \cite{NHRev1,NHRev2,NHRev3,NHRev4,NHRev5,NHRev6}
for reviews). Theoretically, these unique phases have been classified
according to their symmetries and exceptional topologies~\cite{NHClass1,NHClass2,NHClass3,NHClass4,NHClass5,NHClass6}.
Experimentally, non-Hermitian topological phases have been engineered
in a broad range of physical platforms~\cite{NHExp1,NHExp2,NHExp3,NHExp4,NHExp5,NHExp6,NHExp7,NHExp8,NHExp9,NHExp10,NHExp11,NHExp12,NHExp13,NHExp14},
yielding potential applications such as topological lasers \cite{TILZ1,TILZ2,TILZ3}
and high-performance sensors~\cite{NHSens1,NHSens2,NHSens3,NHSens4}.

The localization problem in non-Hermitian systems was introduced early on by 
Hatano and Nelson~\cite{HNM1}, and followed up by a series of related studies~\cite{HNM2,HNM3,HNM4,HNM5}.
In the Hatano-Nelson model, the asymmetry in the hopping amplitude of the one-dimensional
Anderson model was found to be able to induce a transition from an insulator to a metallic phase.
Recently, the development of non-Hermitian topological matter has re-evoked the interest in non-Hermitian systems
with spatial aperiodicity. Specially, the interplay between non-Hermitian
effects and quasiperiodic modulations has been found to induce rich
patterns of localization-delocalization transitions, topological phase
transitions and mobility edges in one-dimensional~(1D) non-Hermitian
quasicrystals~(NHQCs)~\cite{NHQC1,NHQC2,NHQC3,NHQC4,NHQC5,NHQC6,NHQC7,NHQC8,NHQC9,NHQC10,NHQC11,NHQC12,NHQC13,NHQC14,NHQC15,NHQC16,NHQC17,NHQC18,NHQC19,NHQC20,NHQC21,NHQC22,NHQC23,NHQC24,NHQC25}.
Typical systems considered in the study of 1D NHQCs include various
extensions of the Aubry-Andr\'e-Harper~\cite{NHQC1,NHQC2,NHQC3,NHQC4,NHQC5},
Fibonacci~\cite{NHQC17}, Su-Schrieffer-Heeger~\cite{NHQC21}, Kitaev
chain~\cite{NHQC24} and Maryland~\cite{NHQC25} models. Localization
and topological transitions induced by time-periodic driving fields
have also been investigated in the context of Floquet NHQCs~\cite{NHQC26}.
Another way to make a disordered system non-Hermitian is to couple it to the outside world by a lead, as
explored earlier in \cite{OpenQC}.

In this work, we introduce a non-Hermitian extension of the quasiperiodic
Maryland model~\cite{MM1,MM2,MM3,MM4,MM5,MM6} by adding nonreciprocity to its hopping amplitudes.
The resulting system exhibits two distinct NHQC phases
and a phase transition induced by the hopping asymmetry. We introduce
our model in Sec.~\ref{sec:Mod} and reveal its key features in Sec.~\ref{sec:Res}. 
Besides characterizing the real-to-complex spectrum
transition and the delocalization transition from an insulator to a mobility
edge phase with coexisting extended and localized states in Secs.~\ref{subsec:E} and \ref{subsec:IPR}, we
also introduce a topological winding number to describe different
phases and transitions in the nonreciprocal Maryland model in
Sec.~\ref{subsec:WN}. We summarize our results and discuss potential future
work in Sec.~\ref{sec:Sum}.

\section{Model\label{sec:Mod}}
The Maryland model describes particles hopping in a tight-binding
lattice and subject to an unbounded onsite superlattice potential.
Its Hamiltonian in position representation takes the form $H_{0}=\sum_{n}(J|n\rangle\langle n+1|+{\rm H.c.}+V\tan(\pi\alpha n)|n\rangle\langle n|)$.
Here $J$ is the nearest-neighbor~(NN) hopping amplitude, $V$ is the amplitude
of onsite potential, $n\in\mathbb{Z}$ is the lattice site index and
$\{|n\rangle\}$ forms a complete basis of the lattice. $\alpha$
is chosen as an irrational number to realize quasiperiodic
modulations. The Hermitian Maryland model $H_{0}$ is first
introduced as an integrable model to study localization problems in
quantum chaotic systems~\cite{MM1,MM2,MM3,MM4,MM5,MM6}. Later, it
is also utilized to understand the localization in higher dimensions~\cite{MM7,MM8,MM9} 
and topological features of integer quantum Hall
effects~\cite{MM10}. Experimentally, the Maryland model might be
realized in photonic systems by engineering the light propagation
in polygonal optical waveguide lattices \cite{MMExp1}.

In this work, we focus on the localization problem in a non-Hermitian
extension of the Maryland model, which is obtained by incorporating
asymmetry into the hopping amplitudes of $H_0$. The Hamiltonian
of the resulting system, which is dubbed the nonreciprocal Maryland
model (NRMM), takes the following form
\begin{alignat}{1}
H & = J\sum_{n}(e^{-\gamma}|n\rangle\langle n+1|+e^{\gamma}|n+1\rangle\langle n|)\nonumber \\
& + V\sum_{n}\tan(\pi\alpha n)|n\rangle\langle n|.\label{eq:H}
\end{alignat}
Here $J,V,\gamma\in\mathbb{R}$. $\gamma$ measures the degree of
asymmetry between left-to-right and right-to-left NN
hopping amplitudes. We take the periodic boundary condition~(PBC)
for all calculations below by identifying $|n\rangle=|n+L\rangle$,
where $n=1,2,...,L$ and $L$ is the length of lattice. 
When $\alpha=p/q$ ($p,q$ being coprime integers) is chosen to be a rational number and $L$ is an integer multiple of $q$, the system is in the commensurate regime and under the PBC it is expected to hold charge density wave like states. In this work, we instead take
$\alpha=\frac{\sqrt{5}-1}{2}$, i.e., the inverse golden ratio
to yield a quasiperiodic potential. Inserting
the expansion of state $|\psi\rangle=\sum_{n}\psi_{n}|n\rangle$ into
the eigenvalue equation $H|\psi\rangle=E|\psi\rangle$, we obtain
\begin{equation}
J(e^{-\gamma}\psi_{n+1}+e^{\gamma}\psi_{n-1})+V\tan(\pi\alpha n)\psi_{n}=E\psi_{n}.\label{eq:Seq}
\end{equation}
Here $E$ is the energy of state $|\psi\rangle$, which is in general
complex as $H\neq H^{\dagger}$. $\psi_{n}\equiv\langle n|\psi\rangle$
represents the amplitude of right eigenvector $|\psi\rangle$ on the
$n$th lattice site. The solution of Eq.~(\ref{eq:Seq}) under the
PBC then yields all the possible eigenenergies and eigenstates of
the NRMM. Note that in numerical calculations, we take a rational
approximation for $\alpha$ by setting $\alpha\simeq F_{l}/F_{l+1}$,
where $F_{l}$ and $F_{l+1}$ are two adjacent elements of the Fibonacci
sequence. 

In the Hermitian limit ($\gamma=0$), due to the unbounded nature
of onsite potential $V_{n}=V\tan(\pi\alpha n)$, all eigenstates of
$H$ are localized with energy-dependent localization lengths for
irrational $\alpha$ and $V\neq0$. Away from the Hermitian limit,
however, we find a nonreciprocity induced transition of the system
from a localized phase with real spectrum to a mobility edge phase
with complex spectrum at a finite $\gamma=\gamma_{c}$, which is thus
of non-Hermitian origin. 
Note that the mobility edge phase means a phase in which extended and localized
states coexist and are separated in their energies by a mobility edge.
We present systematic characterizations of
this transition and the resulting energy-dependent mobility edges
in the following section. 

\section{Results\label{sec:Res}}
In this section, we investigate the spectrum, delocalization and topological
transitions of the NRMM. In Sec.~\ref{subsec:E}, we study the
energies of NRMM and find a real-to-complex spectral transition
at a finite hopping asymmetry $\gamma=\gamma_{c}$, whose expression as
a function of the hopping amplitude and onsite potential is obtained. 
In Sec.~\ref{subsec:IPR}, the spectrum transition is
connected to a transition of the system from localized to mobility
edge phases. The mobility edge separating localized and extended states
is further picked up and its expression is found to be energy-dependent.
In Sec.~\ref{subsec:WN}, a topological winding number is introduced
to distinguish phases with different transport nature and characterize
the transitions between them in the NRMM, thus yielding a complete
phase diagram. For ease of reference, we summarize
our main results about the NRMM in Table~\ref{tab:NRMM}.
\begin{table*}
	\begin{centering}
		\begin{tabular}{|c|c|c|}
			\hline 
			\textbf{Phase} & \textbf{Localized} & \textbf{Mobility edge}\tabularnewline
			\hline 
			\hline 
			\textbf{Condition} & $|2J\sinh\gamma|<|V|$ & $|2J\sinh\gamma|>|V|$\tabularnewline
			\hline 
			\textbf{Spectrum} & Real & Complex~(see Eq.~(\ref{eq:EPM}))\tabularnewline
			\hline 
			\textbf{IPR} & $>0$ for all states & $>0$ and $\simeq0$ coexist\tabularnewline
			\hline 
			\textbf{Mobility edge equation} & \multicolumn{2}{c|}{$\frac{V^{2}}{(2J\sinh\gamma)^{2}}+\frac{({\rm Re}E)^{2}}{(2J\cosh\gamma)^{2}}=1$}\tabularnewline
			\hline 
			\textbf{Winding number} & \multicolumn{2}{c|}{$w=\int_{0}^{2\pi}\frac{d\theta}{2\pi i}\partial_{\theta}\ln\det[H(\theta)-E_{0}]=\begin{cases}
				0 & {\rm Localized}\\
				\pm1 & {\rm Mobility\,\,edge}
				\end{cases}$}\tabularnewline
			\hline 
		\end{tabular}
		\par\end{centering}
	\caption{Summary of results for the NRMM. $J$, $\gamma$ and $V$ control the hopping amplitude, hopping asymmetry and
		onsite potential. $E$ is the eigenenergy. $H(\theta)$
		is obtained from $H$ in Eq.~(\ref{eq:H}) by setting $e^{\pm\gamma}\rightarrow e^{\pm\gamma\pm i\theta/L}$,
		with $L$ being the length of lattice. The base energy is set as $E_{0}=0$
		in the calculation of $w$.\label{tab:NRMM}}
\end{table*}

\subsection{Real-to-complex spectrum transition\label{subsec:E}}
\begin{figure}
	\begin{centering}
		\includegraphics[scale=0.47]{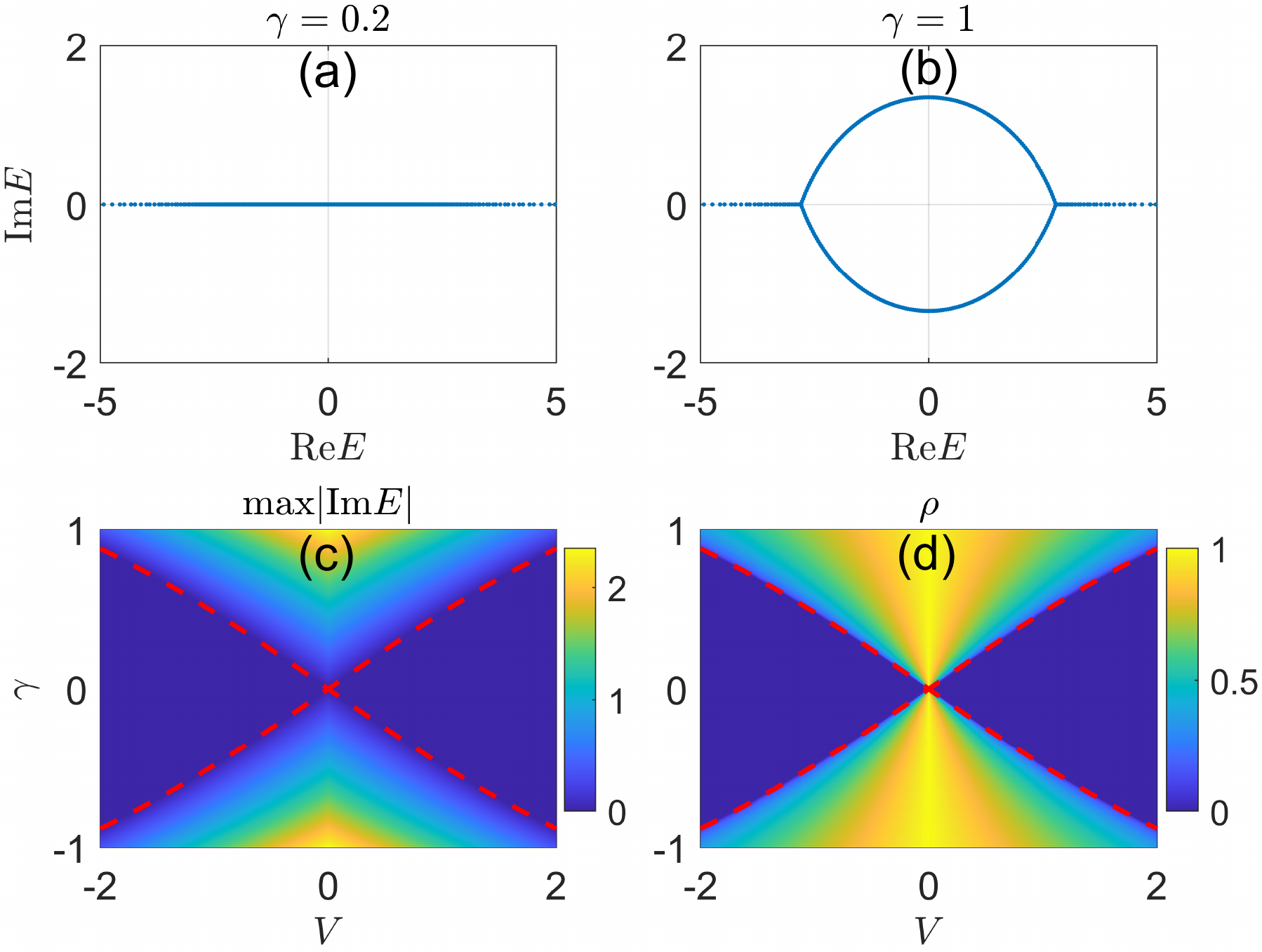}
		\par\end{centering}
	\caption{Spectrum and DOS of NRMM under the PBC. System
		parameters are $J=1$, $\alpha=\frac{\sqrt{5}-1}{2}$ for all
		panels, and the length of lattice is $L=377$. (a) and (b)
		show two typical examples of the spectrum on the complex plane with
		$V=1$. (c) and (d) show the maximal imaginary parts of eigenenergies
		$\max|{\rm Im}E|$ and DOS with complex energies $\rho$ at different
		sets of $(V,\gamma)$. The red dashed lines in (c) and (d)
		highlight the boundaries between phases with real and complex spectrum,
		which are determined by Eq.~(\ref{eq:Gc}).\label{fig:E}}
\end{figure}

We first study the spectrum of NRMM by solving the eigenvalue
Eq.~(\ref{eq:Seq}). Two typical examples of the spectrum are presented
in Figs.~\ref{fig:E}(a) and \ref{fig:E}(b), where ${\rm Re}E$ and
${\rm Im}E$ refer to the real and imaginary parts of energy
$E$, respectively. We observe that with weak hopping asymmetry $\gamma$,
the eigenvalues of $H$ could retain real. When $\gamma$ goes
beyond a critical value $\gamma_{c}$, a finite amounts of eigenenergies
deviate from the real axis, and the spectrum undergoes a
real-to-complex transition. After the transition, the complex part of eigenenergies
develop a loop on the ${\rm Re}E$-${\rm Im}E$ plane surrounding
a base energy $E_{0}=0$. 
The points along the loop in Fig.~\ref{fig:E}(b) satisfy the equation
\begin{equation}
E_{\pm}=2J\cos(\beta-i\gamma)\pm iV,\quad\beta\in[-\pi,\pi)\label{eq:EPM}
\end{equation}
under the constraint ${\rm Im}{E_{-}}>{\rm Im}{E_{+}}$.

To find the critical point $\gamma_{c}$ as a function of system parameters,
we evaluate the maximum of imaginary parts of energy $\max|{\rm Im}E|$
and the density of states~(DOS) with complex eigenvalues $\rho$ at different
$V$ and $\gamma$, with results presented in Figs.~\ref{fig:E}(c)
and \ref{fig:E}(d). In numerical calculations, we count $E$ as a
complex eigenvalue if $|{\rm Im}E|>10^{-5}$. It is clear that once
$V\neq0$, we could obtain spectrum transitions from real to complex
in the NRMM with the increase of $|\gamma|$. By setting
Im$E_{\pm}=0$ in Eq.~(\ref{eq:EPM}), we find 
the critical values of hopping asymmetry $\gamma=\gamma_c$ where spectrum
transitions happen, i.e.,
\begin{equation}
\gamma_{c}=\pm{\rm arcsinh}\left(\frac{V}{2J}\right).\label{eq:Gc}
\end{equation}
When $|\gamma|<|\gamma_{c}|$, all eigenvalues of $H$ are found to
be real, whereas a finite portion of the spectrum becomes complex
for $|\gamma|>|\gamma_{c}|$. We plot these exact phase boundaries
by red dashed lines in Figs. \ref{fig:E}(c) and \ref{fig:E}(d),
and find that they are coincide with numerical calculations
of the spectrum. Meanwhile, Eq.~(\ref{eq:Gc}) provides
us with a guideline for the study of transport nature of the NRMM,
as will be discussed in the next subsection.

\subsection{Delocalization transition and mobility edge\label{subsec:IPR}}
In NHQCs, spectrum transitions usually accompany state
transitions regarding their spatial profiles~\cite{NHQC1,NHQC5}. In the NRMM,
we also discover a transition from an insulator phase
with no extended eigenstates to a mobility edge phase, in which extended
and localized eigenstates coexist. To see this, we first inspect the
inverse participation ratio~(IPR), which is defined for the $i$th
normalized eigenstate $|\psi^{i}\rangle$ of $H$ in the lattice representation as 
${\rm IPR}_{i}={\textstyle \sum_{n=1}^{L}}|\psi_{n}^{i}|^{4}$.
Here the amplitude $\psi_{n}^{i}=\langle n|\psi^{i}\rangle$ and $i=1,2,...,L$.
In the localized phase, all eigenstates have finite IPRs. Extended
states start to appear when the minimum of IPRs, denoted as $\min({\rm IPR})$,
starts to approach zero.

\begin{figure}
	\begin{centering}
		\includegraphics[scale=0.47]{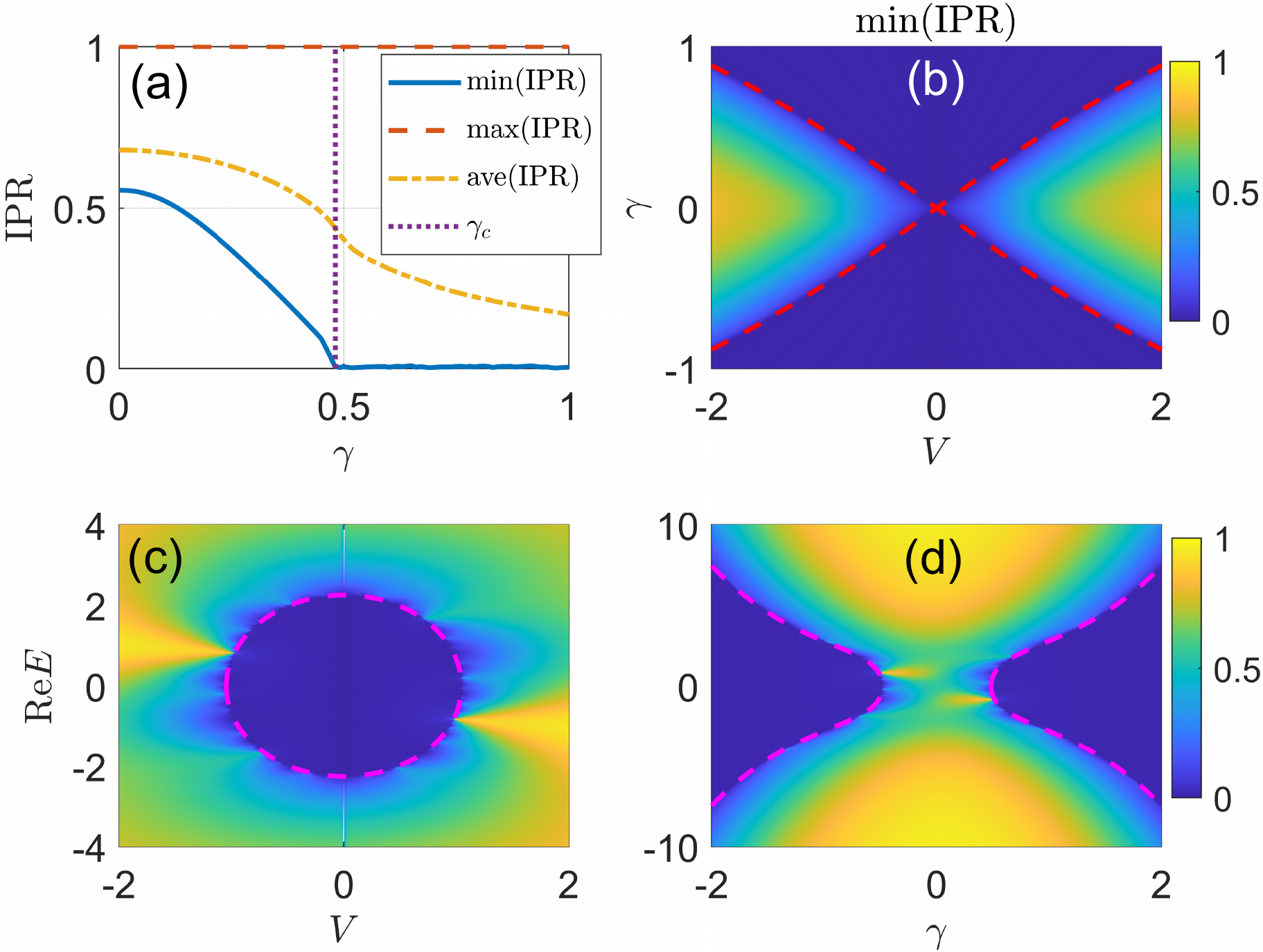}
		\par\end{centering}
	\caption{IPRs of NRMM under the PBC. The length of lattice is 
		$L=987$. System parameters are $J=1$, $\alpha=\frac{\sqrt{5}-1}{2}$
		for all panels. In (a), the solid, dashed and dash-dotted lines
		show the minimum, maximum and average of IPRs versus the hopping asymmetry
		$\gamma$ at $V=1$. The crossing point of the dotted line and the
		horizontal axis corresponds to the critical value of $\gamma=\gamma_{c}={\rm asinh}(V/2J)\approx0.4812$,
		where the system undergoes a transition from the localized to
		the mobility edge phase. (b) shows the minimum of IPRs at different
		parameters $(V,\gamma)$. The red dashed lines refer to the
		boundaries between localized and mobility edge phases, which are given
		by Eq.~(\ref{eq:Gc}). The IPRs of all states versus the real parts
		of their energies and $V$ ($\gamma$) are shown in (c) ((d))
		with $\gamma=0.5$ ($V=1$). The magenta dashed lines obtained following Eq.~(\ref{eq:ME}) are
		mobility edges separating extended and localized states.\label{fig:IPR}}
\end{figure}

In Fig.~\ref{fig:IPR}(a), we show the minimum,
maximum and average of IPRs for the NRMM. It is clear that the $\max({\rm IPR})\simeq1$
due to the unbounded nature of the onsite potential $V_{n}$, which
indicates that there are localized states at any
hopping asymmetry in the limit $L\rightarrow\infty$. Notably, the $\min({\rm IPR})$ decreases
to zero when $\gamma$ goes beyond a critical value $\gamma_{c}$,
which happens to be coincide with the critical point of spectrum transition
in Eq.~(\ref{eq:Gc}). Therefore, when the hopping asymmetry is tuned
from below to above the critical point $\gamma_{c}$, the NRMM transforms
from a localized phase with real spectrum to a mobility edge phase with
complex spectrum. The number of extended states in the mobility edge
phase further increases with the increase of $\gamma$, as hinted
by the ${\rm ave}({\rm IPR})$ in Fig.~\ref{fig:IPR}(a). In Fig.~\ref{fig:IPR}(b), 
we present the $\min({\rm IPR})$ as a function
of system parameters $(V,\gamma)$, and indeed observe two distinct
phases characterized by $\min({\rm IPR})>0$ and $\min({\rm IPR})\simeq0$.
Their boundaries are highlighted by the red dashed lines and
given exactly by Eq.~(\ref{eq:Gc}). To the best of our knowledge,
the NRMM contributes the first example of a 1D NHQC with only
localized and mobility edge phases, but no metallic phases 
at any amounts of non-Hermiticity. 

To give a more detailed look at the mobility edge, we show the IPRs
of all eigenstates of the NRMM versus the real parts of their energies
and the potential amplitude $V$~(hopping asymmetry $\gamma$) for
two typical examples in Figs.~\ref{fig:IPR}(c) and \ref{fig:IPR}(d).
We observe that the states with ${\rm IPR}\simeq0$ and ${\rm IPR}>0$
are clearly separated in both figures. Moreover, with thorough numerical
analysis, we find an equation that describes the mobility edge of
the NRMM, i.e.,
\begin{equation}
\frac{V^{2}}{(2J\sinh\gamma)^{2}}+\frac{({\rm Re}E)^{2}}{(2J\cosh\gamma)^{2}}=1.\label{eq:ME}
\end{equation}
Trajectories determined by this equation, presented by the magenta
dashed lines in Figs.~\ref{fig:IPR}(c) and \ref{fig:IPR}(d), separate states with
vanishing and finite IPRs in the energy-parameter plane of the system.
Moreover, the Eq.~(\ref{eq:ME}) is well-defined at finite $V$ only
if $\gamma\neq0$. Therefore, mobility edges in the NRMM are solely
originated from non-Hermitian effects encoded in the hopping asymmetry
of the lattice.

\subsection{Topological invariant and phase diagram\label{subsec:WN}}
In recent studies, a spectral winding number has been introduced
to depict the transitions between different NHQC
phases~\cite{NHQC1}, following a strategy that is different from the topological characterization
of Hermitian quasicrystals~(see \cite{QCRev1} for a review). 
The general idea is to incorporate a periodic
parameter $\theta$ into the Hamiltonian $H$ of the system, usually
achieved by imposing the twist boundary condition~(flux insertion),
and then track the spectral flow of the parameter-dependent Hamiltonian
$H(\theta)$ with respect to a certain base energy on the complex
plane during the change of $\theta$ over a cycle. If the spectrum
of $H(\theta)$ is real, all eigenvalues collapse onto the real
axis and their winding number $w$ must vanish with respect to any
base energies. If $H(\theta)$ possesses complex energies, their
flow with respect to $\theta$ could form loops around certain points
on the complex plane, and the spectral winding number of $H(\theta)$
around these points could be nonzero and quantized. 
The value of $w$ jumps whenever the spectral of the system changes from real to complex, or vice versa. Therefore, $w$ can be naturally employed as a topological order parameter to distinguish NHQC phases with real and complex eigenspectrum. Interestingly, it was found in some typical NHQC models~\cite{NHQC1,NHQC26} that the real-complex spectral transition could go together with a localization-delocalization transition (with a possible exception reported in Ref.~\cite{NHQC23}). Therefore, we may adopt the same winding number to signify the emergence of mobility edge phase in our system, since the extended states therein are those whose energies possess nonvanishing imaginary parts.

For the NRMM, we construct a spectral winding number as
\begin{equation}
w=\frac{1}{2\pi i}\int_{0}^{2\pi}\partial_{\theta}\ln\det[H(\theta)-E_{0}]d\theta.\label{eq:WN}
\end{equation}
Here $H(\theta)$ is obtained from $H$ by letting $e^{\pm\gamma}\rightarrow e^{\pm\gamma\pm i\theta/L}$
in Eq.~(\ref{eq:H}), with $L$ being the length of lattice. This winding
number counts the number of times the spectrum of $H(\theta)$
winds around the base energy $E_{0}$ when $\theta$ changes over
a cycle from zero to $2\pi$.
Referring to the spectrum presented in Fig.~\ref{fig:E}, we choose
$E_{0}=0$ in our calculation of $w$ without loss of generality.
It is clear that in the Hermitian region~($\gamma=0$) we have $w=0$,
since in this case $H(\theta)$ has only real energies. At a finite
hopping asymmetry $\gamma=\gamma_c$, we expect a quantized jump of $w$
from zero to $\pm1$, which should be accompanied by the spectrum
transition of the NRMM from real to complex. 

\begin{figure}
	\begin{centering}
		\includegraphics[scale=0.48]{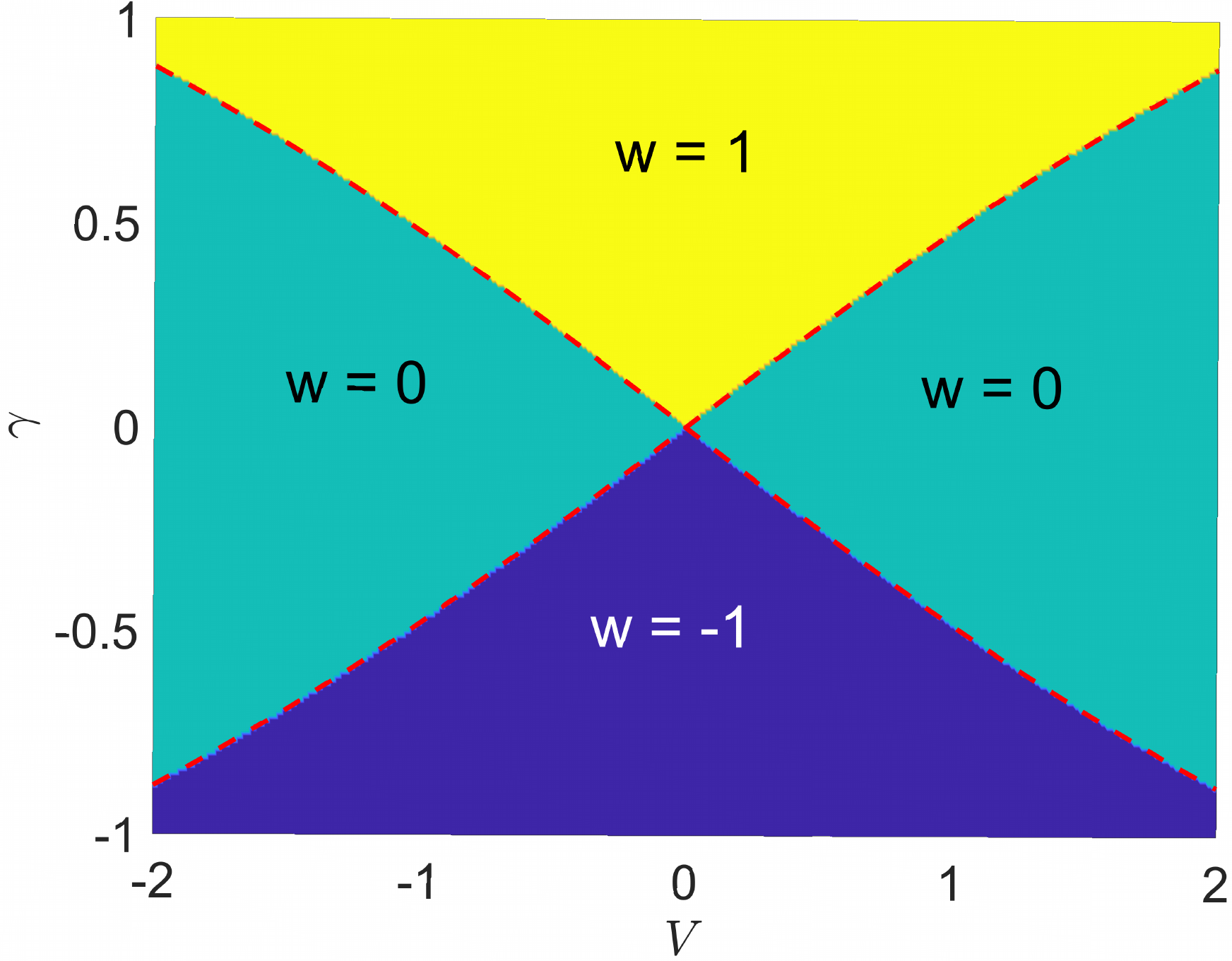}
		\par\end{centering}
	\caption{Winding numbers of the NRMM. System parameters are
		$J=1$ and $\alpha=\frac{\sqrt{5}-1}{2}$. The length of lattice
		is $L=377$ with the twist boundary condition. Each region with a
		uniform color corresponds to a phase with common spectrum and transport
		features, whose winding number $w$ is denoted explicitly therein.
		The red dashed lines separating different regions denote the phase
		boundaries obtained from Eq.~(\ref{eq:Gc}).\label{fig:WN}}
\end{figure}

In Fig.~\ref{fig:WN}, we present the winding number $w$ of NRMM
versus the onsite potential $V$ and hopping nonreciprocity $\gamma$,
which is obtained directly from Eq.~(\ref{eq:WN}). The red dashed
lines in Fig.~\ref{fig:WN} highlight the exact phase boundaries,
which are given by Eq.~(\ref{eq:Gc}). We observe that the winding
number $w$ takes a quantized change when crossing the border between
two NHQC phases, and remain constant elsewhere. More specially, we
find $w=0$ when the system is prepared in the localized phase with
real spectrum, and $w=\pm1$ if the system moves into the mobility
edge phase with complex spectrum. These observations confirm that
the topological winding number $w$ can indeed be utilized to discriminate
the NHQC phases of NRMM with distinct spectrum and transport nature,
and characterize the transitions between them in the meantime.

For completeness, we have checked other values of the irrational parameter $\alpha$ (e.g., $\alpha=(\sqrt{5}+1)/2$, $1/\sqrt{2}$) in our calculations of spectrum, IPRs and winding numbers of the NRMM. The results we obtained are all consistent with those reported in Figs.~\ref{fig:E}--\ref{fig:WN}, which implies that the conclusion we drew from this section is general to other incommensurate values of the onsite potential in Eq.~(\ref{eq:H}).

\section{Summary and discussion\label{sec:Sum}}

In this work, we uncover non-Hermiticity induced spectral, localization
and topological phase transitions in the nonreciprocal Maryland model.
A transformation of the system from a localized phase with real spectrum
to a mobility edge phase with complex spectrum can be obtained only
if there are finite amounts of hopping asymmetry. The equations satisfied
by the complex spectrum, phase boundaries and energy-dependent mobility edges are found
exactly. A topological order parameter is further employed to build
the phase diagram and characterize transitions between different NHQC
phases. Notably, due to the unbounded nature of
onsite potential in the NRMM, we find no extended phases
at any finite hopping nonreciprocity, which is distinct
from the situation found in a Maryland model with complex onsite
potential~\cite{NHQC25}. Our study thus enriches the family of NHQCs
by unveiling a particular type of system with only localized
and mobility edge phases, whose properties can be characterized
exactly. 

In future work, it would be interesting to consider many-body effects in Maryland-type NHQCs and investigate their dynamical properties. The interplay between non-Hermiticity and quantum chaos in the NRMM can further be studied following the mapping discussed in Appendix \ref{app:1}. In this work, our calculations are performed under the PBC and the non-Hermitian skin effects (NHSEs) \cite{NHSE1,NHSE2,NHSE3,NHSE4,NHSE5} are expected to have no impacts on the results. Under the open boundary condition, the nonreciprocal hoppings in our model could possibly induce non-Hermitian skin modes. Recently, it was shown that (de)localization transitions and (pseudo) mobility edges could even emerge in clean systems due to the NHSEs, and topological characterizations of these intriguing phenomena have been proposed~\cite{PMBZhou}. Therefore, it is expected that in the presence of both NHSEs and spatial quasiperiodicity, richer patterns of spectrum, localization and topological transitions could appear in generic NHQCs, which deserve more thorough explorations.

\begin{acknowledgments}
L.Z. is supported by the National Natural Science Foundation of China~(Grant No.~11905211), 
the China Postdoctoral Science Foundation~(Grant
No.~2019M662444), the Fundamental Research Funds for the Central Universities~(Grant No.~841912009), the Young Talents Project at Ocean University
of China~(Grant No.~861801013196), and the Applied Research Project
of Postdoctoral Fellows in Qingdao~(Grant No.~861905040009).
\end{acknowledgments}

\appendix

\section{NRMM and Floquet system}\label{app:1}
\setcounter{equation}{0}
\setcounter{figure}{0}
\numberwithin{equation}{section}
\numberwithin{figure}{section}

The Hermitian Maryland model can be mapped to a mathematically equivalent
Floquet system~\cite{MM1}. A similar mapping can also be constructed for
the NRMM, which may serve as an entrance for the study of the interplay
between non-Hermiticity and quantum chaos.

We start by rewriting the eigenvalue Eq.~(\ref{eq:Seq}) as
\begin{equation}
\frac{E}{V}\psi_{n}-\frac{e^{-\gamma}}{V}\psi_{n+1}-\frac{e^{\gamma}}{V}\psi_{n-1}=\tan(\pi\alpha n)\psi_{n},\label{eq:Seq1}
\end{equation}
where we have set $J=1$ as the unit of energy. Multiplying the imaginary
unit $i$ from both sides of Eq.~(\ref{eq:Seq1}), we obtain
\begin{equation}
\frac{V\psi_{n}-i(E\psi_{n}-e^{-\gamma}\psi_{n+1}-e^{\gamma}\psi_{n-1})}{V\psi_{n}+i(E\psi_{n}-e^{-\gamma}\psi_{n+1}-e^{\gamma}\psi_{n-1})}=e^{-i2\pi\alpha n}.\label{eq:Seq2}
\end{equation}
For a 1D quasicrystal, the amplitude $\psi_{n}$ can be expressed
as a superposition of plane waves~\cite{NHQC25}, i.e.,
\begin{equation}
\psi_{n}=\sum_{\ell}\varphi_{\ell}e^{ik_{\ell}n}.\label{eq:psin}
\end{equation}
Here for any $\ell\neq\ell'$, the difference between wave numbers
$k_{\ell}$ and $k_{\ell'}$ is an integer multiple of $2\pi\alpha$.
Taking $\psi_{n}$ as amplitudes, we can further construct a series
\begin{alignat}{1}
\Psi(x)& = \sum_{n}\psi_{n}e^{inx}=\sum_{\ell}\varphi_{\ell}\sum_{n}e^{i(x+k_{\ell})n}\nonumber \\
& = 2\pi\sum_{\ell,n}\varphi_{\ell}\delta(x+k_{\ell}-2\pi n),\label{eq:PsiX}
\end{alignat}
where we used the Poisson summation formula to arrive at the last
equality. If we now multiply $e^{inx}$ to Eq.~(\ref{eq:Seq2}) and
take the summation over $n$, we will obtain with the help of Eq.~(\ref{eq:PsiX}) that
\begin{alignat}{1}
& \left[1-i\frac{E}{V}+i\frac{2}{V}\cos(x-i\gamma)\right]\Psi(x)\label{eq:Seq3}\\
= & \left[1+i\frac{E}{V}-i\frac{2}{V}\cos(x-2\pi\alpha-i\gamma)\right]\Psi(x-2\pi\alpha).\nonumber 
\end{alignat}

To make the connection between the NRMM and its Floquet equivalent
more transparent, we introduce the function
\begin{equation}
\Phi(x)=\left[1+i\frac{E}{V}-i\frac{2}{V}\cos(x-i\gamma)\right]\Psi(x).\label{eq:PhiX}
\end{equation}
Using Eq.~(\ref{eq:PhiX}), we can express Eq.~(\ref{eq:Seq3}) as
\begin{equation}
\Phi(x)=\frac{1+i\frac{E}{V}-i\frac{2}{V}\cos(x-i\gamma)}{1-i\frac{E}{V}+i\frac{2}{V}\cos(x-i\gamma)}\Phi(x-2\pi\alpha).\label{eq:PhiX2}
\end{equation}
Performing the Taylor expansion of $\Phi(x-2\pi\alpha)$, we find
\begin{equation}
\Phi(x-2\pi\alpha)=\sum_{n}\frac{(-2\pi\alpha)^{n}}{n!}\frac{\partial^{n}\Phi(x)}{\partial x^{n}}=e^{-2\pi\alpha\partial_x}\Phi(x).\label{eq:PhiX3}
\end{equation}
Meanwhile, we can introduce a function ${\cal K}(x)$ that satisfies
\begin{alignat}{1}
e^{-i{\cal K}(x)}& = \cos[{\cal K}(x)]-i\sin[{\cal K}(x)]\nonumber \\
 &= \frac{1+i\frac{E}{V}-i\frac{2}{V}\cos(x-i\gamma)}{1-i\frac{E}{V}+i\frac{2}{V}\cos(x-i\gamma)},\label{eq:EiKx}
\end{alignat}
yielding
\begin{equation}
{\cal K}(x)=2\arctan\left[\frac{2}{V}\cos(x-i\gamma)-\frac{E}{V}\right].\label{eq:Kx}
\end{equation}
Plugging Eqs.~(\ref{eq:PhiX3})--(\ref{eq:Kx}) into Eq.~(\ref{eq:PhiX2}),
we finally obtain
\begin{equation}
\Phi(x)=e^{-i{\cal K}(x)}e^{-2\pi\alpha\partial_{x}}\Phi(x),\label{eq:PhiX4}
\end{equation}
which can be interpreted as describing the one-period evolution of
a particle with linear dispersion in its kinetic energy $-2\pi\alpha i\partial_{x}$
and subject to a delta-kicking potential ${\cal K}(x)$ within the
driving period $T=1$. The corresponding Schr\"odinger equation takes
the form
\begin{equation}
i\partial_{t}\Phi=-2\pi\alpha i\partial_{x}\Phi+{\cal K}(x)\sum_{m\in{\mathbb Z}}\delta(t-m)\Phi,\label{eq:SeqK}
\end{equation}
and the quasienergy can be viewed as $\varepsilon=0$. The above analysis
establishes a relationship between the spectrum nature of NRMM
and the dynamics of a periodically kicked particle with linearized
kinetic energy.

\end{document}